\documentclass[prd,column,superscriptaddress,amsfonts,amssymb,amsmath,showpacs]{revtex4-2}
\usepackage{bm}
\usepackage{latexsym}
\usepackage[latin1]{inputenc}
\usepackage{graphicx}
\usepackage{amsmath}
\usepackage{amssymb}
\usepackage{amsthm}
\usepackage{relsize}
\usepackage{palatino}
\usepackage{mathpazo}
\usepackage{textcomp}
\usepackage{float}
\usepackage{booktabs}
\usepackage{dcolumn}
\usepackage{booktabs}
\usepackage{multirow}
\usepackage{tikz}
\usepackage{hyperref}
\hypersetup{
    colorlinks=true,
    linkcolor=purple,
    filecolor=magenta,      
    citecolor=blue
}
\usepackage{amsmath}
\usepackage{xcolor}
\usepackage{orcidlink}
\usepackage[caption=false]{subfig}
\usepackage{commath}
\makeatletter
\long\def\dddddot#1{%
  {\mathop {#1}\limits ^{\vbox to-1.4\ex@ {\kern -\tw@ \ex@ \hbox {\normalfont .....}\vss }}}%
}
\long\def\multidots#1#2{%
  \count@=0
  {{\mathop {#2}\limits ^{\vbox to-1.4\ex@ {\kern -\tw@ \ex@ \hbox {\normalfont %
  \loop%
  \ifnum#1>\count@%
  .%
  \advance\count@ by1%
  \repeat%
  }\vss }}}}%
}
\makeatother

\begin{document}

\title{\bf Observational constrained Weyl type $f(Q,T)$ gravity cosmological model and the dynamical system analysis }

\author{Rahul Bhagat\orcidlink{0009-0001-9783-9317}}
\email{rahulbhagat0994@gmail.com}
\affiliation{Department of Mathematics, Birla Institute of Technology and
Science-Pilani,\\ Hyderabad Campus, Hyderabad-500078, India.}
\author{B. Mishra\orcidlink{0000-0001-5527-3565}}
\email{bivu@hyderabad.bits-pilani.ac.in}
\affiliation{Department of Mathematics, Birla Institute of Technology and
Science-Pilani,\\ Hyderabad Campus, Hyderabad-500078, India.}

\begin{abstract}
Using the cosmological date sets, the cosmological parameters are constrained in this paper, with some well known form of Hubble parameter. To understand the dynamics of the Weyl type $f(Q,T)$, functional form $f(Q,T)$ has been introduced, where $Q$ and $T$ respectively represents the nonmetricity scalar and trace of energy-momentum tensor. Using the constrained values of the parameters, the other geometrical parameters are analysed and the accelerating behaviour has been shown. Further to get the complete evolutionary behaviour of the Universe, the dynamical system analysis has been performed.
\end{abstract}

\maketitle

\section{Introduction} 
Revealing the accelerating expansion behaviour of the Universe \cite{Riess_1998_116, Perlmutter_1998_517} is a remarkable finding in modern cosmology. For a better understanding of the cause of such behaviour, the scientific community have focused on the ways to measure and describe it. This has led to the establishment of large-scale facilities that can generate extensive datasets for analysis; and the cosmological probes have emerged as the standard techniques. A number of cosmological probes have been thoroughly studied over time \cite{Cole_2005_362, Daniel_2008_77, Moresco_2022_25}. These probes include Baryon Acoustic Oscillations (BAO) \cite{Bennett_2003_148, Eisenstein_2005_633}, Cosmic Microwave Background (CMB)\cite{Larson_2011_192, Komatsu_2011_192} and Supernovae Type Ia (SNe) \cite{Ade_2016_594a, Abbott_2016_460}. The large-scale structure study has revealed the existence of correlated over--densities in matter distribution at a specific separation of $r \sim 100 Mpc/h$, known as Baryon Acoustic Oscillations \cite{Eisenstein_2005_633, Hinshaw_2013_208}. Type Ia Supernovae has been identified as particularly useful standard candles at cosmological distances because their peak luminosity correlates precisely with their absolute luminosity following appropriate calibration\cite{Phillips_1993_413}. Concurrently, examining the initial light emitted into the Universe , known as CMBR through various ground and space missions has offered valuable insights into the early Universe \cite{Balkenhol_2023_108}. Other cosmological investigation has been extensively employed in recent decades to delineate the expansion of the Universe and the progression of matter within it \cite{Carlstrom_2011_123}.\\

It has been indicated that some form of exotic energy, known as dark energy (DE) is responsible for this behaviour of the Universe and the need for a negative pressure indicates the non-baryonic nature of DE.  To understand the effect of DE in general relativity (GR), the most straightforward and thoroughly investigated assumption is to include a cosmological constant factor \cite{Fujii_1982_26}. The cosmological constant can be interpreted as a vacuum energy component in quantum field theory, which leads to a negative equation of state (EoS) parameter.  At an equivalent model, the $\Lambda$CDM, can be readily obtained by using the cosmological constant. However, if someone modifies the geometric part of  Einstein-Hilbert action, it can be performed in three approaches, such as curvature, torsion and nonmetricity. To note when one of the approaches is considered the other two remains zero. We shall use here the nonmetricity approach to modify the geometrical part of Einstein-Hilbert action \cite{Nester_1999}. The corresponding gravitational theory known as the symmetric teleparallel, where the basic geometry of the gravitational action is represented by the nonmetricity $Q$. Further the symmetric teleparallel gravity developed into the coincident GR or $f(Q)$ gravity \cite{Jimenez_2018_98}. The cosmological and astrophysical aspects of $f(Q)$ gravity has been successful to some extents. Anagnostopoulos et al. \cite{Anagnostopoulos_2021_822} proposed a model that maintains the same number of free parameters as in $\Lambda$CDM; but unlike $\Lambda$CDM, the model diverges from the limits of $\Lambda$CDM. Expressing $f(Q)$ Lagrangian explicitly as the function of redshift, Lazkoz et al. \cite{Lazkoz_2019_100} observational constrained the parameters.  Capozziello et al. \cite{Capozziello_2022_832} introduced a method for reconstructing the $f(Q)$ action without resorting $priori$ assumptions about the cosmological model. Using dynamical system analysis and perturbation techniques, Khyllep et al. \cite{Khyllep_2021_103} investigated the background cosmological scenario. Paliathanasis \cite{Paliathanasis_2023_41} explored the evolution of physical variables in $f(Q)$-gravity for two families of symmetric and flat connections within a spatially flat FLRW geometry. Narawade et al.\cite{Narawade_2023_535} obtained the Hubble parameter as a function of redshift through algebraic manipulation of the $f(Q)$ form considered, and they examined the accelerated expansion of the Universe using observational datasets.\\

With the introduction of trace of energy momentum tensor $T$, the $f(Q)$ gravity further extended to $f(Q,T)$ gravity \cite{Xu_2019_79}. N\'ajera et al. \cite{Najera_2021_34,Najera_2022_2022} developed a linear cosmological perturbations and proposed five specific models, each with parameters leading to a $\Lambda$CDM limit. Through analysis using cosmic chronometers and supernovae Ia data, they demonstrated that this model aligns with $\Lambda$CDM with a confidence level of 95\%. Further, Narawade et al. \cite{Narawade_2023_992} presented an accelerating cosmological model by utilizing a particular parametric form of the Hubble parameter and conducting a dynamical system analysis to confirm the stability of the model. Bhagat et al. \cite{Bhagat_2023_42} explored the cosmological acceleration of the Universe by deriving constraints on free parameters within the non-linear form of nonmetricity in $f(Q,T)$ through the construction of a parametrized Hubble parameter. Another development in the nonmetricity based gravitational theory is the Weyl type $f(Q,T)$ gravity \cite{Xu_2020_80}, which is the extension of the $f(Q)$ \cite{Nester_1999} type family of gravity. In this proposal, a vector field $\omega_\mu$ completely determines the scalar nonmetricity $Q_{\alpha\nu\mu}$ of the space-time, which  can be expressed in conventional Weyl form $(Q=-6\omega^2)$. The field equations of Weyl type $f(Q,T)$ gravity are derived on the basis of the assumption that the total scalar curvature vanishes. The gravitational Lagrangian in Weyl type $f(Q,T)$ gravity creates dependency on the nonmetricity scalar $Q$ and trace of energy-momentum tensor $T$. One of the noteworthy characteristics of the Weyl type $f(Q,T)$ gravity is that it violates the conservation of energy principle in its standard form. This is because there can be an energy exchange between matter and curvature in the Lagrangian due to the non-minimal interaction between these two elements. Here, the gravitational action was varying about both metric and connection to develop the field equations. Without requiring the presence of DE, a comprehensive theoretical explanation of the late-time cosmic acceleration of the Universe can be obtained through the matter-energy coupling in this $f(Q,T)$ gravity. Some research has been made in this gravity to examine the dynamical features of the expanding Universe \cite{Bhagat_2023_41}. Yang et al. reviewed the evolution of kinematical quantities associated with deformations in the Weyl type $f(Q,T)$ gravity, considering the geodesic deviation equation and the Raychaudhuri equation \cite{Yang_2021_81}. Additionally, Koussour et al. used observational data to investigate the compatibility of exact solutions in Weyl-type $f(Q,T)$ gravity \cite{Koussour_2023_83}.\\

The paper has been organized as follow: In Sec. \ref{Sec:2}, the field equations of Weyl type $f(Q,T)$ gravity has been discussed. In Sec. \ref{Sec:3}, using the cosmological observations such as $CC$, $Pantheon^+$ and $BAO$ datasets, the cosmological  parameters are constrained. In Sec.\ref{Sec:4}, the dynamical behaviour of the cosmological model has been analysed with the constrained value of the parameters. Further, the dynamical system analysis has been performed for the model in Sec. \ref{Sec:5}. Finally, we summarise our results and conclusion in Sec. \ref{Sec:6}.

\section{Field Equations of Weyl type $f(Q,T)$ gravity}\label{Sec:2}

The Weyl type $f(Q,T)$ gravity action \cite{Xu_2020_80} is given as,
\begin{eqnarray}\label{eq:1}
S = \int \sqrt{-g} \big[\frac{1}{16\pi G} f(Q,T) -\frac{1}{4} W_{\mu\nu} W^{\mu\nu} -\frac{1}{2}m^{2} \omega_{\mu} \omega^{\mu}
+\lambda (R +6\bigtriangledown_\alpha \omega^ \alpha-6 \omega_\alpha \omega^ \alpha) +\mathcal{L}_m \big]dx^4 ~,
\end{eqnarray}
 where $W_{\mu\nu}=\triangledown_\mu \omega_\nu-\triangledown_\nu \omega_\mu$ in the form of Weyl vector, $\mathcal{L}_m$ be the matter Lagrangian and $m$ be the mass of the particle corresponding to the vector field. The second term $\left[\frac{1}{4} W_{\mu\nu} W^{\mu\nu}\right]$ represents the standard kinetic term whereas the third term  $\left[\frac{1}{2}m^{2} \omega_{\mu} \omega^{\mu}\right]$ represents the mass term of the vector field. The semi-metric connection in Weyl geometry is given as,
\begin{equation}
     \tilde \Gamma ^\lambda_{\mu\nu}=\Gamma^{\lambda}_{\mu\nu}+g_{\mu\nu}w^\lambda-\delta_\mu^\lambda w_\nu - \delta_\nu^\lambda w_\mu.
\end{equation}
The Christoffel symbol, $\Gamma ^\lambda_{\mu\nu}$ constructed with respect to the metric $g_{\mu\nu} $. The nonmetricity tensor is,

\begin{align}\label{eq:2}
 Q_{\alpha\mu\nu}&\equiv\tilde\triangledown_\alpha g_{\mu\nu}=\partial_\alpha g_{\mu\nu} -\tilde \Gamma ^\rho _{\alpha \mu}g_{\rho \nu}-\tilde \Gamma ^\rho_{\alpha\nu}g_{\rho\mu}\nonumber\\& = 2\omega_\alpha g_{\mu\nu},
\end{align}
 The scalar nonmetricity is,
\begin{equation}\label{eq:3}
Q \equiv -g^{\mu\nu} (L^\alpha_{\beta\mu}L^\beta _{\nu\alpha} - L^\alpha _{\beta\alpha} L^\beta _{\mu \nu}),
\end{equation}
with
\begin{equation}\label{eq:4}
L^\lambda _{\mu\nu}=-\frac{1}{2}g^{~\lambda\gamma}(Q_{~\mu\gamma\nu}+Q_{~\nu\gamma\mu}-Q_{~\gamma\mu\nu}),
    \end{equation}
Substituting Eq.\eqref{eq:2} in Eq.\eqref{eq:4}, one can obtain
\begin{equation}\label{eq:5}
Q=-6 \omega ^2.
\end{equation}
Varying the action with respect to the vector field, the generalized Proca equation can be obtained as,

\begin{equation}
    \bigtriangledown^{\nu}W_{\mu\nu}-(m^2+12\kappa^2~f_Q+12\lambda)\omega_\mu=6\bigtriangledown_{\mu}\lambda.
\end{equation}
Applying the variation principle on Eq. \eqref{eq:1}, the field equation of Weyl type $f(Q, T)$ gravity can be obtained as ,
\begin{eqnarray}\label{eq:6}
\frac{1}{2}(T_{\mu\nu}+S_{\mu\nu})-\kappa^2 f_T (T_{\mu\nu}+\Theta_{\mu\nu})=-\frac{\kappa^2}{2}g_{\mu\nu}f-6\kappa^2 f_Q \omega_\mu \omega_\nu+\lambda(R_{\mu\nu}-6\omega_\mu \omega_\nu+3g_{\mu\nu}\triangledown_\rho \omega^\rho)\\+3g_{\mu\nu}\omega^\rho\triangledown_\rho\lambda-6\omega_{(\mu}Q_{\nu)}\lambda+g_{\mu\nu}\square \lambda- \triangledown_\mu\triangledown_\nu \lambda,
\end{eqnarray}
where $\kappa^2=\frac{1}{16\pi G}$. The derivative of the function $f(Q,T)$ with respect to $Q$ and $T$ can be denoted respectively as $f_Q$ and $f_T$. Now,
\begin{equation}\label{eq:7}
T_{\mu\nu}\equiv -\frac{2}{\sqrt{-g}}\frac{\delta(\sqrt{-g}\mathcal{L}_m)}{\delta g^{\mu\nu}},
\end{equation}
\begin{equation}\label{eq:9}
\Theta_{\mu\nu} \equiv g^{\alpha\xi}\frac{\delta T_{\alpha \beta}}{\delta g_{\mu\nu}}=g_{\mu\nu}\mathcal{L}_m-2T_{\mu\nu}-2g^{\alpha\beta}\frac{\delta^2\mathcal{L}_m}{\delta g^{\mu\nu}\delta g^{\alpha\beta}}.
\end{equation}\\
The rescaled energy momentum tensor $S_{\mu\nu}$ in the field equation is,
\begin{equation}\label{eq:10}
S_{\mu\nu}=-\frac{1}{4}g_{\mu\nu}W_{\rho\sigma}W^{\rho\sigma}+W_{\mu\rho}W^\rho_\nu-\frac{1}{2}m^2 g_{\mu\nu}\omega_\rho \omega^\rho+m^2\omega_\mu \omega_\nu~.
\end{equation}
To frame the cosmological model of the Universe, we consider the background geometry as flat FLRW space time with the uniform spatial expansion $a(t)$ as,
\begin{equation}\label{eq:11}
ds^2=-dt^2+a^2(t)[dx^2+dy^2+dz^2].
\end{equation}
The Hubble parameter $H=\frac{\dot a}{a}$ with an over dot represents derivative with respect to $t$. Because of the spatial symmetry, the vector field is in the form,
\begin{equation}\label{eq:12}
\omega_\nu=[\psi(t),~0,~0,~0]
\end{equation}
So, $\omega^2=\omega_\nu\omega^\nu=-\psi^2(t)$, $Q=6\psi^2(t)$. The comoving coordinate system has been used, so $u^\mu=(-1,~0,~0,~0)$ and hence $u^\mu\triangledown_\mu=\frac{d}{dt}$. We consider the Lagrangian of the perfect fluid, $\mathcal{L}_m=p$. As a result, $T_\nu^\mu=diag(-\rho,~p,~p,~p)$ and $\Theta^\mu_\nu=\delta^\mu_\nu p-2T^\mu_\nu=diag(2\rho+p,-p,-p,-p)$. The generalized Proca equation, which shows the evolution of the Weyl vector, is given as
\begin{eqnarray}\label{eq:13.1}
    \dot{\psi} &=& \dot{H} + 2H^{2} + \psi^{2} -3H\psi~,\nonumber\\
    \dot{\lambda} &=& -\frac{1}{6}~m^{2}_{eff}~\psi ~,\nonumber\\
    \partial_{i}\lambda &=& 0~,
\end{eqnarray}
where $m^{2}_{eff}=m^2+12\kappa^2~f_Q+12\lambda$ is the effective dynamical mass of the vector field. Now, the field equations of Weyl type $f(Q,T)$ gravity is obtained from Eq. \eqref{eq:6} as,
\begin{equation}\label{eq:14}
\kappa^2f_T(\rho+p)+\frac{1}{2}\rho=\frac{\kappa^2}{2}f-\left(6\kappa^2f_Q+\frac{1}{4}m^2\right)\psi^2 -3\lambda(\psi^2-H^2)-3\dot\lambda(\psi-H)
\end{equation}
\begin{equation}\label{eq:15}
-\frac{1}{2}p=\frac{\kappa^2}{2}f+\frac{m^2\psi^2}{4}+\lambda(3\psi^2+3H^2+2\dot H) +(3\psi+2H)\dot\lambda+\ddot\lambda   
\end{equation}
Using Eq. \eqref{eq:13.1}, derivative of $\lambda$ can be eliminated and subsequently,
\begin{equation}\label{eq:19}
    \frac{1}{2}(1+2\kappa^2 f_T)\rho+\kappa^2f_T p=\frac{\kappa^2}{2}f+\frac{m^2\psi^2}{4}+3\lambda(H^2+\psi^2)-\frac{1}{2}m^{2}_{eff}H\psi,
\end{equation}
\begin{equation}\label{eq:20}
    \frac{1}{2}(1+2\kappa^2f_T)(\rho+p)=\frac{m^{2}_{eff}}{6}(\dot{\psi}+\psi^2-H\psi)+2\kappa^2\dot{f_Q}\psi-2\lambda\dot{H}
\end{equation}
 Further substituting $\dot{\psi}$ in Eq. \eqref{eq:20}, we obtain 
\begin{equation}
    \frac{1}{2}(1+2\kappa^2f_T)(\rho+p)=-2\lambda(1-\frac{m^{2}_{eff}}{12\lambda})\dot{H}+\frac{m^{2}_{eff}}{3}(H^2+\psi^2-2H\psi)+2\kappa^2\dot{f_Q}\psi.
\end{equation}
The equation for the energy balance is,
\begin{equation}
    \dot{\rho}+3H(\rho+p)=\frac{1}{1+2\kappa^2f_{T}}(2\kappa^2(\rho+p)\dot{f_T}-f_T(\dot{\rho}-\dot{p})).
\end{equation}

So, the generalized Friedmann Eq. \eqref{eq:19} and Eq. \eqref{eq:20} can be reformulated in an effective form as,
\begin{eqnarray}
3H^2&=&\frac{1}{2\lambda}(\rho+\rho_{eff}),\label{eq:23}\\
3H^2+2\dot{H}&=&-\frac{1}{2\lambda}(p+p_{eff}) \label{eq:24}
\end{eqnarray}
where
\begin{eqnarray}
\rho_{eff}&=&m^{2}_{eff}H\psi+2\kappa^2f_T(\rho+p)-\kappa^2f-\frac{{m^2}{\psi^2}}{2}-6\lambda\psi^2,\label{eq:23a}\\
p_{eff}&=&\frac{m^{2}_{eff}}{3}(\dot{\psi}+\psi^2-4H\psi)+\kappa^2f+4\kappa^2\dot{f_Q}\psi  +\frac{m^2 \psi^2}{2}+6\lambda\psi^2.\label{eq:24a}
\end{eqnarray}
The gravitational action simplifies to the conventional Hilbert-Einstein form when the parameters approach their limiting values. Specifically, when $f=0$, $\psi=0$ and $\lambda=\kappa^2$, both the effective energy density $\rho_{eff}$ and effective pressure $p_{eff}$ are zero and the above field equations reduce to that of GR.\\

\section{Observational Analysis}\label{Sec:3}
We shall use the cosmological data sets to constrain the parameters involved and hence we consider the Hubble parameter \cite{Sahni_2003_77} as,
\begin{equation}
    H(z)=H_0 \sqrt{\alpha(z+1)^4+\Omega_m (z+1)^3+\zeta(z+1)^\frac{1}{6}},
\end{equation}
where $\alpha$, $\zeta$, and $\Omega_m$ are the parameters for this model and $H_0$ denotes the present value of Hubble parameter. The deceleration parameter can now be,

\begin{equation}\label{Eqn:26}
q(z)=\frac{12~\alpha~(z+1)^4+6~\Omega_m~(z+1)^3-11~\zeta~\sqrt[6]{z+1}}{12 \left(\alpha~(z+1)^4+\Omega_m~(z+1)^3+\zeta~\sqrt[6]{z+1}\right)}
\end{equation}
First we shall use the following cosmological data sets to constrain the parameters involved in the parameterization.

{\bf{Cosmic Chronometer:}}  We have used the $32$ Hubble data points, measured in the redshift range $0.07\leq z \leq 1.965$, using the differential age (DA) method \cite{Jimenez_2002_573}. It allows to determine the Universe expansion rate at redshift $z$. Hence, $H(z)$ can be determined using $H(z)=-\frac{1}{1+z}\frac{dz}{dt}$. To obtain the mean values of the model parameters $H_0$, $\alpha$, $\Omega_m$ and $\zeta$, the chi-square for Hubble datasets can be expressed as, 
\begin{equation}\label{eq.18}
\chi^{2}_{CC}(z,H_0,\alpha,\Omega_m,\zeta)=\mathlarger{\mathlarger{\sum}}_{i=1}^{32}\frac{\big[H_{th}(z_{i},H_0,\alpha,\Omega_m,\zeta)-H_{obs}(z_i)\big]^2}{\sigma_{CC}^2(z_i)},
\end{equation}

where $H_{th}(z_{i},H_0,\alpha,\Omega_m,\zeta)$ denote the theoretical value of the Hubble parameter, $H_{obs}(z_i)$ denotes the observed value of the Hubble parameter and the standard error is denoted as $\sigma_{CC}^2(z_i)$. We shall use 32 CC sample \cite{Moresco_2022_25}.\\

{\bf{Pantheon$^+$ Data Points:}} The $Pantheon^{+}$ dataset includes $1701$ light curves of $1550$ distinct Type Ia supernovae (SNe Ia), in the redshift range  $0.00122\leq z \leq 2.2613$  \cite{Brout_2022_938}. The Universe appears to be expanding at an accelerated rate based on measurements of type Ia supernovae. The distance modulus used to calculate the distance from the supernovae. The model parameters to be fitted by comparing the observed and hypothesised values of the distance moduli. The $\chi^2_{Pantheon^{+}}$ has been given as,
 \begin{equation}\label{eq:22}
 \chi^2_{Pantheon^{+}}(z,H_0,\alpha,\Omega_m,\zeta)=\sum_{i=1}^{1701}\frac{[\mu_{th}(z_i,H_0,\alpha,\Omega_m,\zeta)-\mu(z_i)_{obs}]^2}{\sigma^2_\mu(z_i)},
 \end{equation}
 where $\sigma^2_{\mu}(z_i)$ denotes the standard error in the observed value and $\mu_{th}(z_i,H_0,\alpha,\Omega_m,\zeta)$ be the theoretical distance modulus, which can be written as,
 \begin{equation}\label{eq:23}
 \mu_{th}=\mu(D_L)=m-\mathcal{M}=5\log_{10}D_L(z)+\mu_0,
 \end{equation}
$\mu_0$ is the nuisance parameter and $D_L(z)$ is the dimensionless luminosity distance defined as,
\begin{equation}\label{eq:24}
D_L(z)=(1+z)\int_{0}^{z}\frac{1}{E(\zeta^*)}d\zeta^*.
\end{equation}
$E(z)=\frac{H(z)}{H_0}$  be the dimensionless parameter and $\zeta^*$ denotes the change of variable defined between  $0$ and $z$.\\

{\bf {BAO Dataset}}: It has 6-data points derived from the results of the 6dFGS, SDSS, and Wiggle Z surveys with different redshifts. The measurable quantities can be obtained by using, 

\begin{eqnarray}
d_{A}(z_{*}) = \int_{0}^{z_{*}}\frac{d\tilde{z}}{H(\tilde{z})}, \label{eq:25}\\
D_{V}(z) = \left[\frac{(d_{A}(z))^{2}z}{H(z)}\right]^{\frac{1}{3}}. \label{eq:26}
\end{eqnarray}
$d_{A}(z)$ and $D_{V}(z)$ are respectively denotes the comoving angular diameters and the dilation scale. The chi-square for BAO datasets can be given as,
\begin{equation}\label{eq:27}
\chi _{BAO}^{2}=X^{T}C^{-1}X.
\end{equation}
with $X$ depends on the survey considered and the inverse of covariance matrix $C$ Refs.\cite{Percival_2010_401, Blake_2011_418, Giostri_2012_2012_027}.\\

\begin{figure}[H]
\centering
\includegraphics[width=08cm,height=08cm]{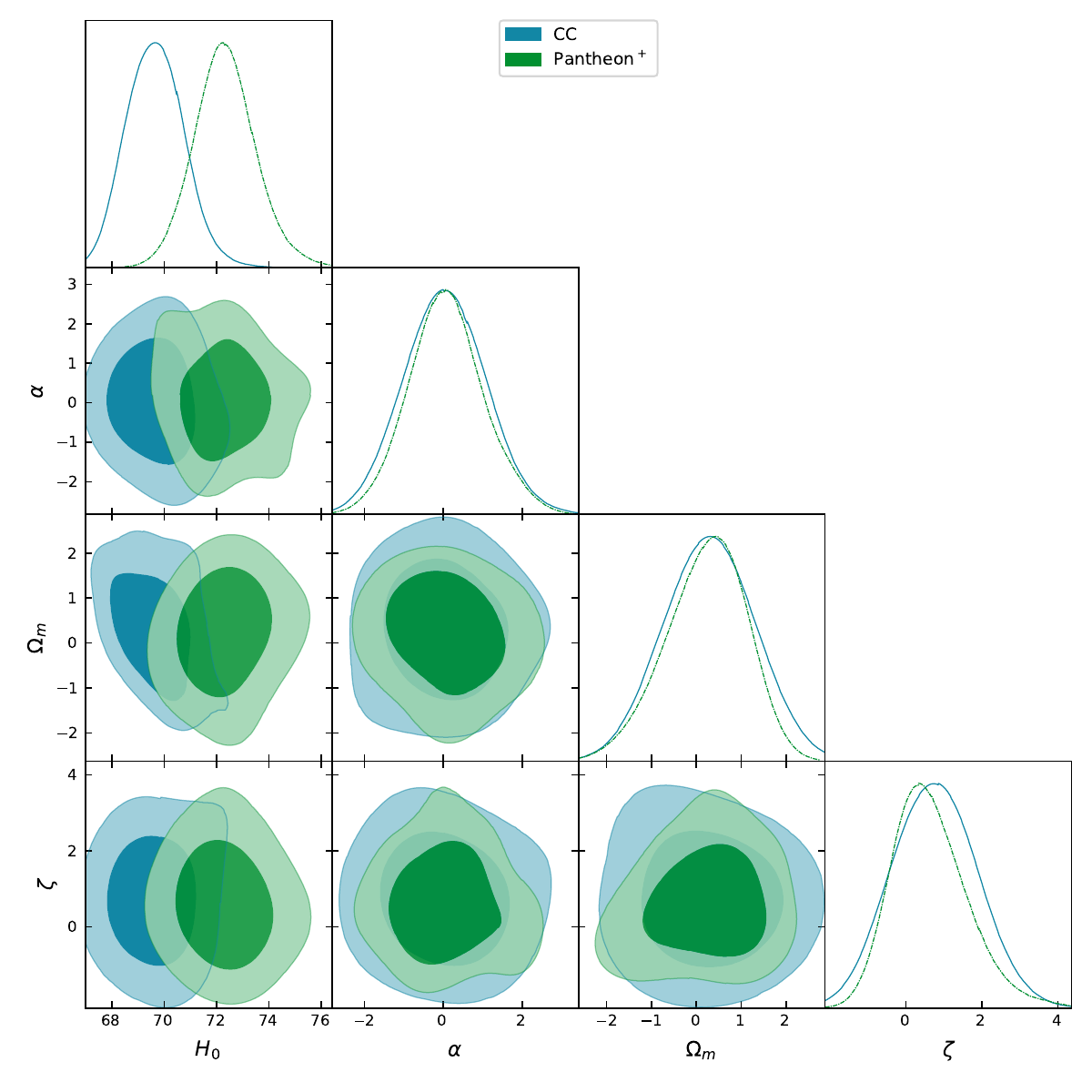}
\includegraphics[width=08cm,height=08cm]{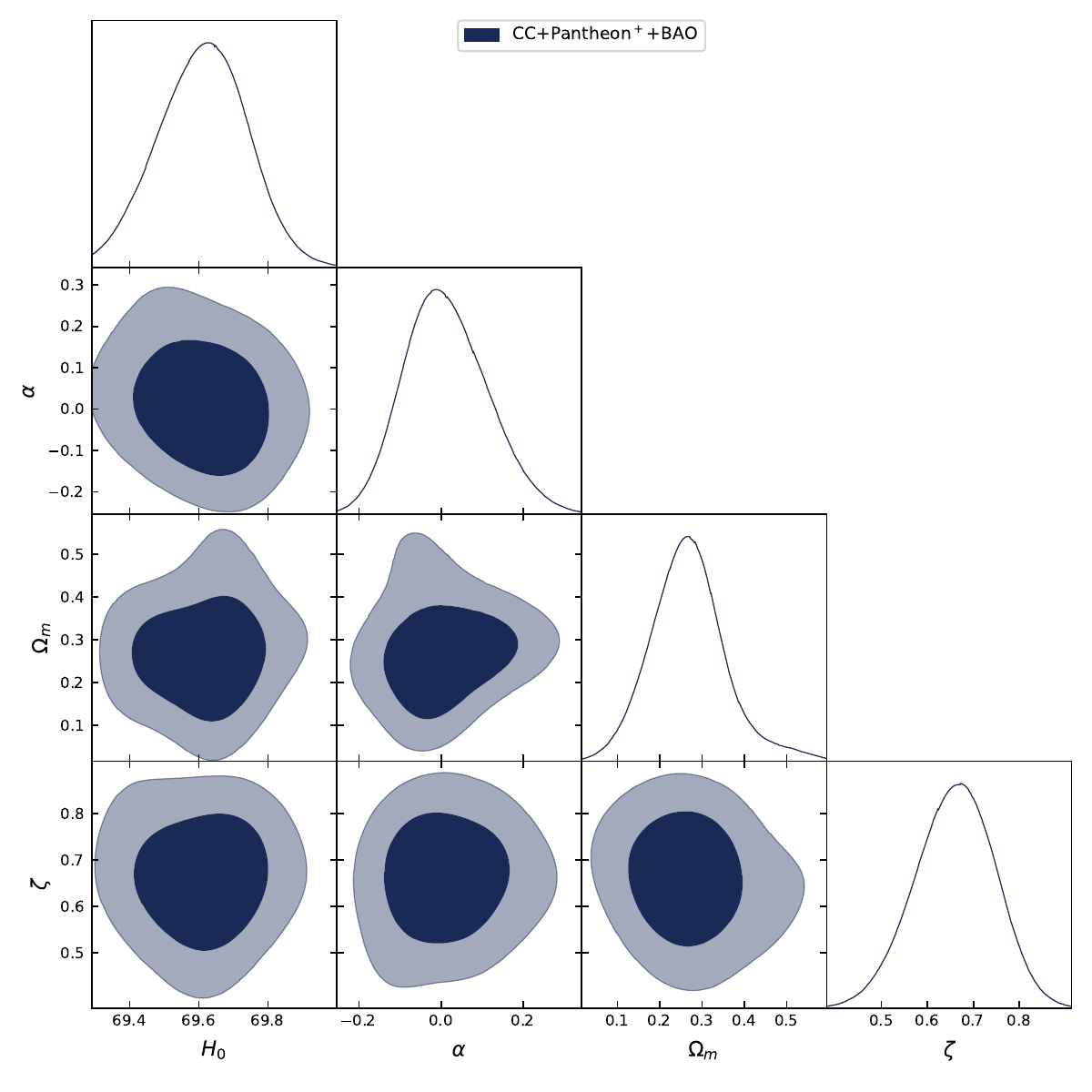}
\caption{The contour plots $CC$, $Pantheon^+$ (\textbf{Left panel}) and $CC+Pantheon^++BAO$ (\textbf{Right panel}) for the parameters $H_{0}$, $\alpha$, $\Omega_m$ and $\zeta$ with $1-\sigma$ and $2-\sigma$ confidence intervals.}
\label{fig:I}
\end{figure}

To explore the parameter space for the likelihood minimization, the MCMC technique \cite{Foreman-Mackey_2013_125} in the python emcee program has been often used. The likelihood contours in FIG.- \ref{fig:I} show how each of the parameters correlates with one another and provide insights into the model degeneracy. Performing the MCMC analysis by using the above mentioned cosmological datasets, the contour plots are drawn [FIG.-\ref{fig:I}]. The constrained values of the parameters obtained from these plots are listed in TABLE-\ref{Table-I} for the individual sample and combined samples.

\begin{table}[H]
\renewcommand\arraystretch{1.5}
\centering 
\begin{tabular}{|c|c|c|c|} 
\hline\hline 
~~~Coefficients~~~& ~~~\textit{CC} Sample  ~~~& ~~~~\textit{Pantheon$^+$}~~~~ & ~~~\textit{CC + Pantheon$^+$ + BAO}~~~\\ [0.5ex] 
\hline\hline
$H_{0}$ & 69.6 $\pm$ 1.0 &  $72.3\pm1.1$ & $69.61 \pm 0.12$ \\
\hline
$\alpha$ & 0.03 $\pm$ 0.99 &  $0.07\pm0.92$ & $0.011^{+0.092}_{-0.11}$ \\
\hline
$\Omega_m$ & $0.29\pm0.93$ &  $0.22^{+0.99}_{-0.79}$ & $0.267^{+0.077}_{-0.089}$ \\
\hline
$\zeta$ & 0.7 $\pm$ 1.0 &  $0.62^{+0.82}_{-1.1}$ & $0.660^{+0.088}_{-0.078}$ \\[0.5ex] 
\hline 
\end{tabular}
\caption{Constrained values of the parameters using $CC$, $Pantheon^+$, $BAO$ datasets.} \label{Table-I}
\label{table:I} 
\end{table}

\begin{figure}[H]
\centering
\includegraphics[scale=0.5]{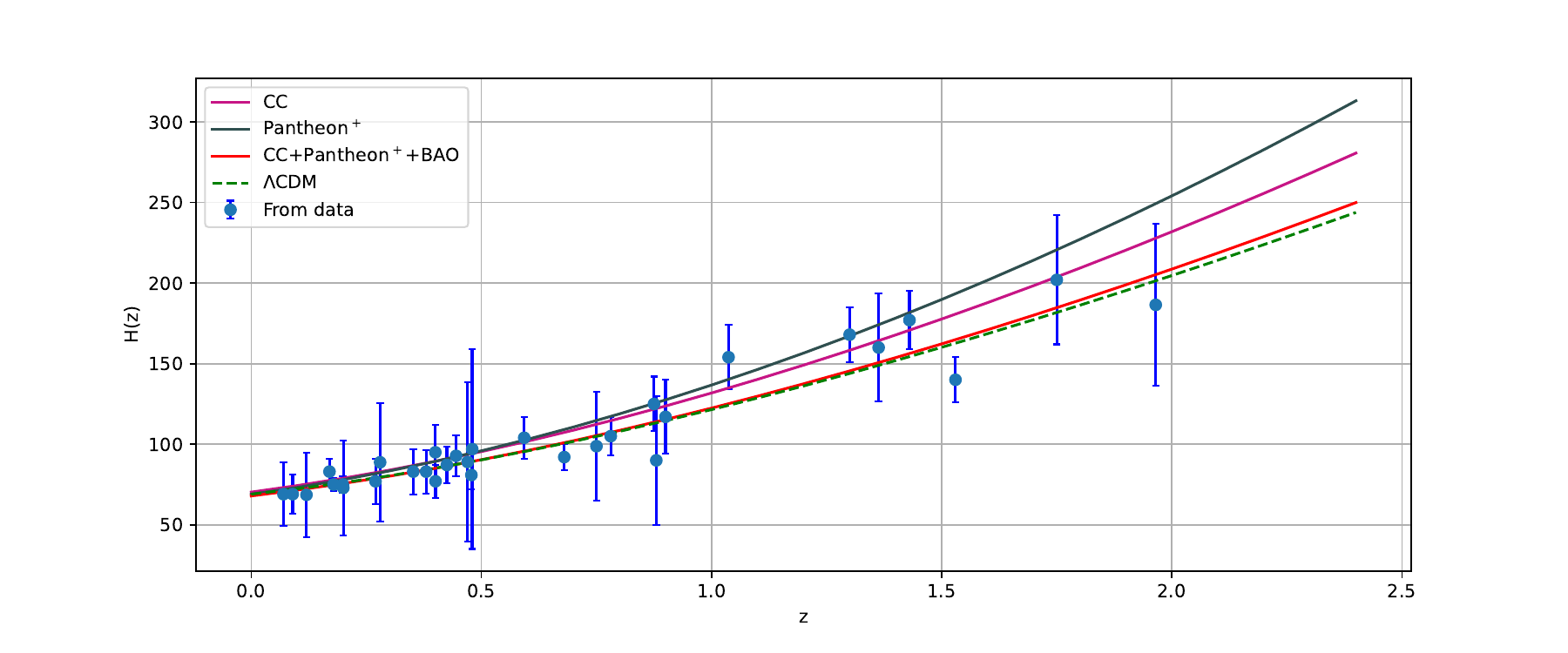}
\includegraphics[scale=0.5]{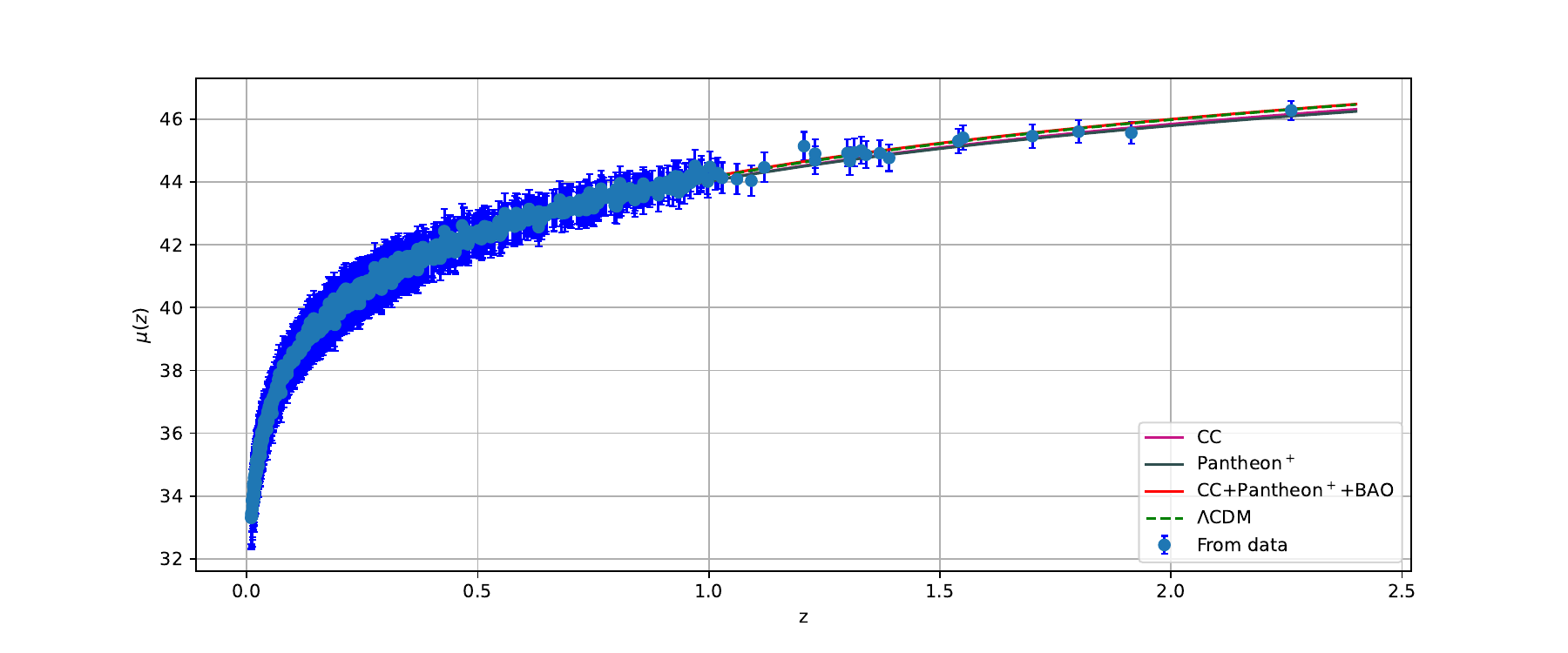}
\caption{Error bar for 32-points of $CC$ sample (\textbf{Upper panel}) and 1701-points of $Supernova$ dataset (\textbf{Lower panel}). The respective curve represents the best-fit values obtained through different data sets.}
\label{fig:II}
\end{figure}

In FIG.\ref{fig:II}, the error plots for the Hubble parameter and the distance modulus measurements are given. These error bars are essential for evaluating the precision of the model predictions since they represent the uncertainty in the observations. The study presented here strongly supports the underlying cosmological model and our current understanding of the expansion history of the Universe. One can observe that the model curve well within the error bars. The $\Lambda$CDM line also shown to observe the deviation from the model curve. This parameterization approach has been widely used to reproduce the cosmological models that shows the accelerating behaviour \cite{Lohakare_2023_39, Agrawal_2023_38_Ob}. The appropriate $\chi^2$ values have been derived using observations from CC, Pantheon$^+$, and BAO datasets. The present value of the expansion of the Universe ($H_0$) for the different data sets, along with the best-fit values for the parameters of $H(z)$, is shown in TABLE \ref{table:I} and FIG.\ref{fig:I}, with $1-\sigma$ and $2-\sigma$ confidence levels. Some cosmological observations suggested the present value of the Hubble parameter as, $H_{0}=72.1\pm 2.0 kms^{-1}Mpc^{-1}$ \cite{Soltis_2021_908}, $H_{0}=69.8\pm 0.6 kms^{-1}Mpc^{-1}$ \cite{Freedman_2021_919} and $H_{0}=68.3\pm 1.5 kms^{-1}Mpc^{-1}$ \cite{Balkenhol_2023_108}. 

\begin{figure}[H]
\centering
\includegraphics[scale=0.5]{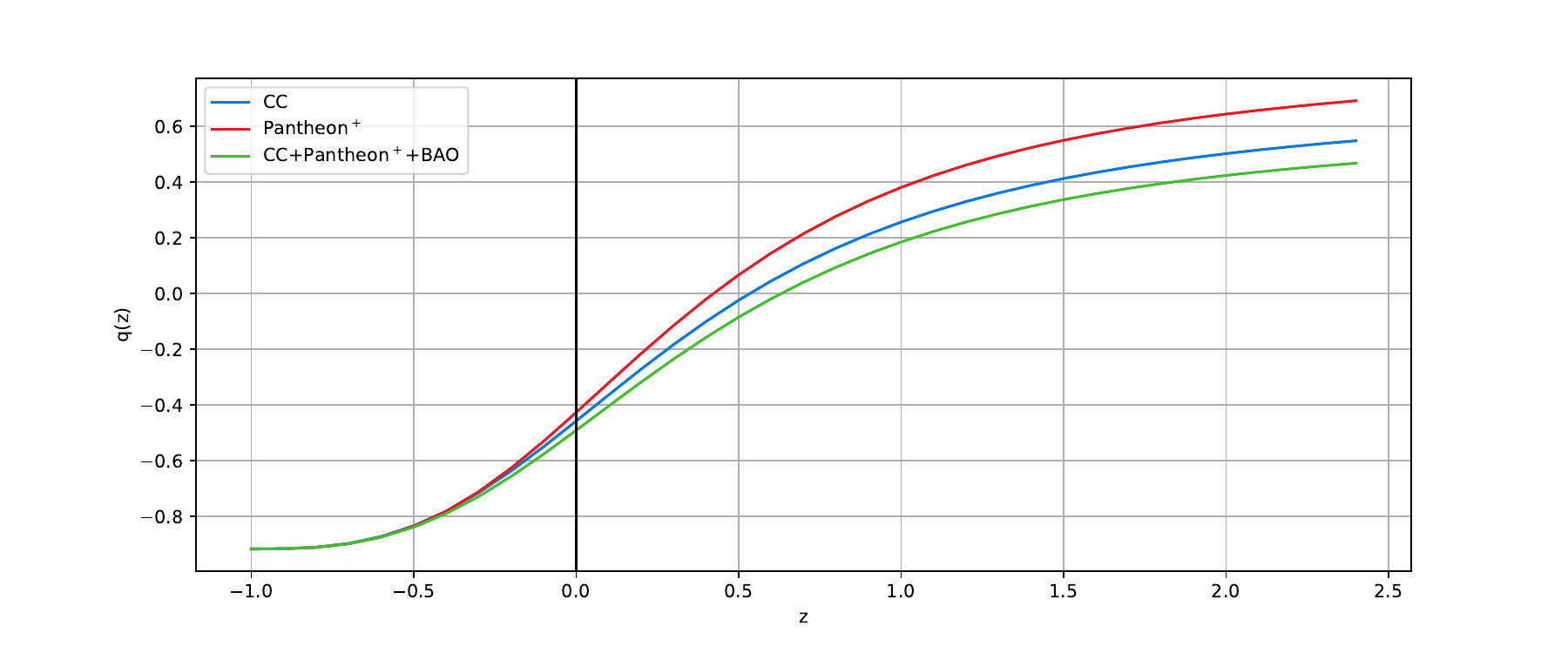}
\caption{Behaviour of deceleration parameter in redshift.}
\label{fig:III}
\end{figure}

The accelerated expansion of the Universe is a late time phenomena and the deceleration parameter as given in Eq. \eqref{Eqn:26} will provide us its behaviour. Now with the observational constrained parametric values, the behaviour of deceleration parameter has been shown in FIG.\ref{fig:III} for the CC, Pantheon$^+$, and BAO datasets. We can observe the transient behaviour of the deceleration parameter from early deceleration to late acceleration. The transition occurred at redshift $z_{tr}=0.541$ for $CC$, $z_{tr}=0.424$ for $Pantheon^+$, and $z_{tr}=0.640$ for the $CC+Pantheon^++BAO$ dataset. Also, the present value of the deceleration parameter are obtained to be $q_0=-0.457$, $q_0=-0.426$, and $q_0=-0.490$ for $CC$, $Pantheon^+$, and $CC+Pantheon^++BAO$ datasets respectively.

\section{Dynamics of Weyl type $f(Q, T)$ Model}\label{Sec:4}
To understand the dynamics of the cosmological model, we need to have some functional form of $f(Q,T)$, which we have considered as $f(Q, T)= \gamma Q + \xi T$ \cite{Xu_2020_80} with $\gamma$ and $\xi$ are model parameters. Now, the effective energy density [Eq. \eqref{eq:23a}] and  effective pressure [Eq. \eqref{eq:24a}]  reduce to, 

\begin{equation}\label{eq.34}
   \rho_{eff}=\frac{1}{2}~H_0^2~((\alpha(1+z)^4+\Omega_m(1+z)^3+\zeta(1+z)^\frac{1}{6})(7+6\gamma)+(1+z)^3\xi~(9~\Omega_m+8(1+z)~\Omega_r)),
\end{equation}
\begin{equation}\label{eq.35}
    p_{eff}=-\frac{1}{36}~H_0^2((42~\alpha~(1+z)^4+36~\Omega_m~(1+z)^3+19~\zeta~(1+z)^\frac{1}{6})(7+6~\gamma)+54~(1+z)^3~\xi~\Omega_m)
\end{equation}
Subsequently, we can obtain the effective EoS parameter as,
 \begin{equation}\label{eq.36}
     \omega_{eff}=\frac{p_{eff}}{\rho_{eff}}=-\frac{(42~\alpha~(1+z)^4+36~\Omega_m~(1+z)^3+19~\zeta~(1+z)^\frac{1}{6})(7+6~\gamma)+54~(1+z)^3~\xi~\Omega_m}{18((\alpha(1+z)^4+\Omega_m(1+z)^3+\zeta(1+z)^\frac{1}{6})(7+6\gamma)+(1+z)^3\xi~(9~\Omega_m+8(1+z)~\Omega_r))}
 \end{equation}

\begin{figure}[H]
\centering
\includegraphics[scale=0.5]{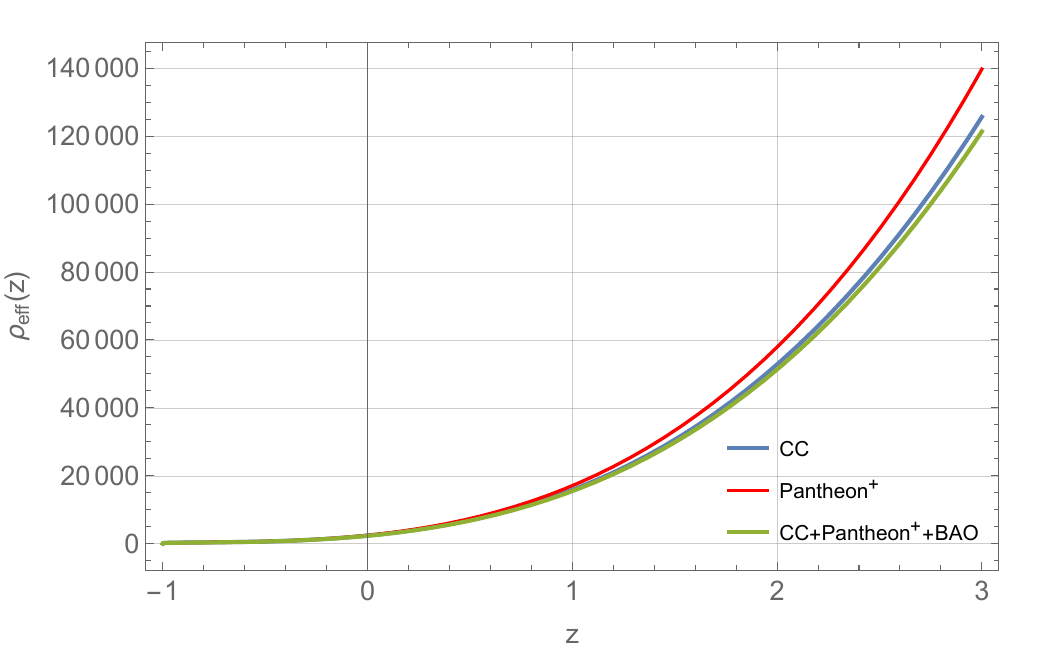}
\includegraphics[scale=0.6]{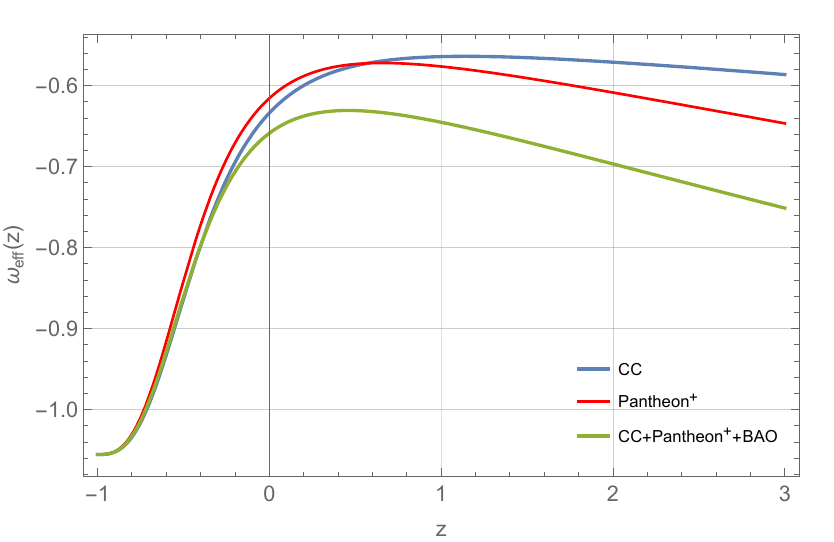}
\caption{Behaviour of energy density and EoS parameter in redshift.}
\label{fig:IV}
\end{figure}

The graphical representation of effective energy density and EoS parameter in redshift has been shown in FIG.\ref{fig:IV} with the constrained values of the parameters obtained from three different datasets. The  effective energy density remains positive throughout and reduces from early to late time. At the same time, the EoS parameter that plays an important role in addressing the late time accelerating phenomena decreases over positive to negative redshift. We have referred to the representative values taken for the parameters $\lambda=\kappa^2=1$ \cite{Bhagat_2023_41} and the value of $\psi$ as $H$ \cite{Xu_2020_80}. The other model parameters representative values are chosen as, $\gamma=-1.12$ and $\xi=0.34$, the effective EoS parameter  shows the quintessence $(-1 < \omega_{eff} < -\frac{1}{3})$ behaviour at present time with the value $\omega_0=-0.633$ for the $CC$ dataset, $\omega_0=-0.615$ for the $pantheon^+$ and $\omega_0=-0.658$ for the $CC+pantheon^++BAO$ data. However at late time, all the curves merging together and show phantom behaviour.

\begin{figure}[H]
\centering
\includegraphics[scale=0.41]{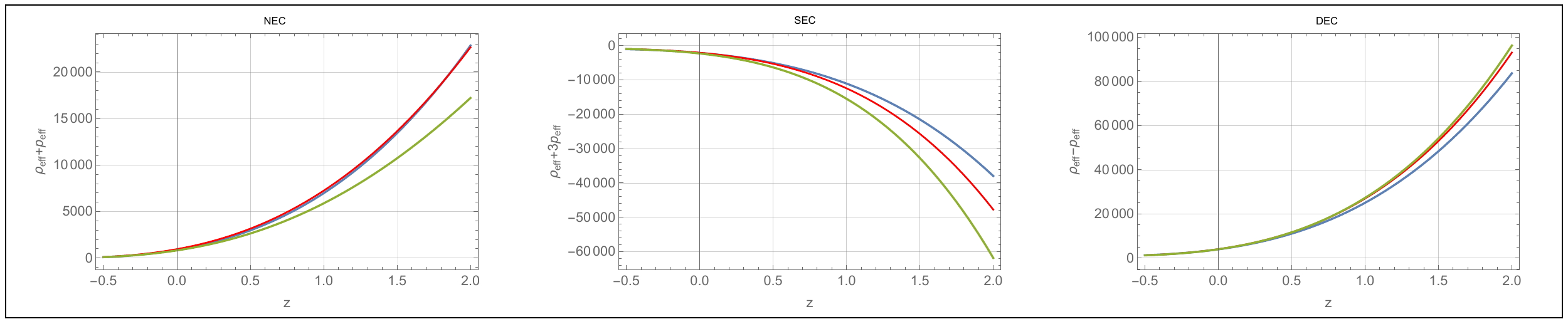}
\caption{Behaviour of Energy conditions in redshift.}
\label{fig:V}
\end{figure}

We study energy conditions that assigns the fundamental causal and geodesic structure of the space-time \cite{Capozziello_2018_781_Erg}. The energy conditions are the boundary conditions to depict the behaviour of energy density in the Universe. The energy conditions are: Null Energy Condition (NEC): $\rho_{eff}+p_{eff} \geq 0$; Dominant Energy Condition (DEC): $\rho_{eff}-p_{eff} \geq 0$; Strong Energy Condition (SEC): $\rho_{eff}+3p_{eff} \geq 0$. The violation of SEC has been inevitable in the context of modified gravity; however NEC may or may not violate. With the considered model parameter values and the observational constrained parameters, we have shown the graphical behaviour of the energy conditions FIG.\ref{fig:V}. One can observe the violation of SEC and the non-violation of NEC and DEC.

\section{The Dynamical system Analysis}\label{Sec:5}
There are several limitations involved while framing the cosmological models to understand the evolutionary behaviour of the Universe. So, to have a comprehensive understanding of the evolution of the Universe for an extended period, the study of dynamic system analysis provides deeper insight \cite{Bohmer_2016_book_dyna}.  Bahamonde et al. \cite{Bahamonde_2019_100_1906.00027} have investigated the dynamics system analysis using the energy-momentum squared gravity model and studied the phase space structure and the physical implications of the energy-momentum squared coupling. Odintsov et al. \cite{Odintsov_2018_98} have shown the occurrence of future cosmological finite-time singularities in the dynamical system corresponding to two cosmological theories. Pati et al. \cite{Pati_2023_83} performed a dynamical system analysis for the hyperbolic scale factor model in $f(Q,T)$ gravity and studied the evolutionary behaviors. Kadam et al. \cite{Kadam_2022_82} consider two different model settings in the context of the dynamical system to reveal their evolutionary behavior. The dynamical system analysis has been performed for scalar field theories to obtain the cosmological behavior of models presented in \cite{Duchaniya_2023_83} with some functional form of the torsion scalar.  

Through this analysis, one may obtain the critical points for the different phases of the Universe. A stable critical point leads to the understanding of stable behaviour of the cosmological model. Here, we shall examine the stability of the systems by reformulating the cosmological equations inside the framework of dynamical system. 

We consider the following dimensionless variables:
\begin{equation}
x=\frac{\Psi}{H},\ \ \ y=\frac{f}{6H^2},\ \ \,z=\frac{\dot H}{H^2},\ \ \,v=\frac{\rho_r}{3H^2}.\ \ \ 
\end{equation}
The Universe is filled with dust and radiation fluids, so that 
\begin{equation}
    \rho = \rho_m + \rho_r, \ \ \ \ \ p_r=\frac{1}{3}\rho_r,\ \ \ \ p_m=0.
\end{equation}
The density parameters for matter, radiation and dark energy are respectively,
\begin{equation}
\Omega_m=\frac{8\pi G \rho_m}{3 H^2},~~~\Omega_r=\frac{8\pi G \rho_r}{3 H^2},~~~\Omega_{de}=\frac{8\pi G \rho_{de}}{3 H^2}
\end{equation}
such that
\begin{eqnarray}
\Omega_m+\Omega_r+\Omega_{de}=1
\end{eqnarray}

Further, the effective EoS and total EoS respectively,
\begin{equation}
\omega_{eff}=\frac{p_{eff}}{\rho_{eff}},
\end{equation}
and the total (EoS) parameter
\begin{equation}
\omega_{tot}=\frac{p_m+p_r+p_{eff}}{\rho_m+\rho_r+\rho_{eff}}\equiv -1-\frac{2\dot{H}}{3H^2}
\end{equation}
In this formalism, the general form of the autonomous dynamical system can be formed by finding the differentiation of the dimensionless variables with respect to $N = ln(a)$ with prime (') denotes the differentiation with respect to the e-fold number $N$. Now,
\begin{align}\label{Eqn.43}
x'&=(x-1) (x-z-2)\,,\nonumber\\
y'&=\frac{\xi  \left(-2 (4 \xi +3) v+7 x^2-2 (6 \gamma +7) x+6 (y+1)\right)}{8 \xi +4}+2 \gamma  x ((x-3) x+z+2)-2 y z\,,\nonumber\\
z'&=\lambda -2 z^2\,,\nonumber\\
v'&=-2 v \left(\frac{6 (\xi +1)}{5 \xi +3}+z\right)\,.
\end{align}
We can also now express the deceleration parameter and effective EoS parameter in terms of dynamical variables as, 
\begin{align}
q&=-1-\frac{\dot{H}}{H^2}\,,\nonumber\\
\omega_{eff}&=\frac{6 (\xi +1) \left(\frac{7 x^2}{6}+\frac{1}{9} (6 \gamma +7) (x (2 x-7)+z+2)+y\right)}{2 \xi  (v+3)+x (12 \gamma -7 x+14)-6 y}\,,\nonumber\\
\omega_{tot}&=-1-\frac{2}{3}z\,.
\end{align}
Subsequently, the density parameters become,
\begin{eqnarray}
\Omega_m&=&\frac{-8 \xi  v-6 v+7 x^2-12 \gamma  x-14 x+6 y+6}{6 \xi +6}\\
\Omega_r&=&v\\
\Omega_{de}&=&\frac{2 \xi  (v+3)+x (12 \gamma -7 x+14)-6 y}{6 (\xi +1)} 
\end{eqnarray}

Now, to determine the critical points from the system Eq. \eqref{Eqn.43} in different evolutionary epoch, we shall search for the points at which $x'=0$, $y'=0$, $z'=0$ and $v'=0$. The stability criteria based on the eigenvalues of the critical points are as follows \cite{Bohmer_2016_book_dyna}:  (a) if all the eigenvalues are real and negative, then the critical point is a stable node or attractor; however if all the eigenvalues are positive, it is an unstable node. (b) If all eigenvalues are real and at least two have opposite signs, it denotes a saddle (unstable) point; (c) if the eigenvalues are complex, then again (i) if all eigenvalues have negative real parts, then a stable focus node; whereas (ii) if all eigenvalues have positive real parts, then an unstable focus node; and (iii) if at least two eigenvalues have real parts with opposite signs, then it is a saddle focus. In the following TABLE--\ref{table2}, we have listed the obtained critical points and its existence condition(s), the eigenvalues and the stability behaviour of each critical points along with its condition(s). The value of the deceleration parameter, EoS parameters and density parameters are listed in TABLE--\ref{table3} and in TABLE--\ref{table4}, the acceleration equation, expansion factor, phase of the Universe and the type expansion for each critical points are listed.

\begin{table}[H]
    \renewcommand\arraystretch{0.5}
    \centering 
    \begin{tabular}{|c| c c c c| c | c c c c |   m{15em}|} 
    \hline\hline 
    \parbox[c][1.3cm]{1.3cm}{Critical Points
    }& $x_c$ & $y_c$ & $z_c$ & $v_c$ & Exits for & $\nu_1$ & $\nu_2$ & $\nu_3$ & $\nu_4$ & \begin{center}
         Stability Conditions
    \end{center}  \\ [0.3ex] 
    \hline\hline 
    \parbox[c][1.3cm]{0.7cm}{$A$ } & $0$ & $0$ & $-2$ & $v_1$ & $\xi_1=0$, $\lambda_1=8$ & $8$ & $4$ & $-1$ & $0$ & \begin{center}
        Unstable
    \end{center}\\
     \hline
    \parbox[c][1.3cm]{0.7cm}{$B$ } & $1$ & $\frac{-1}{6}$ & $z_1$ & $0$ &  \begin{tabular}{@{}c@{}}$\gamma_2=\frac{-1}{6}$, $\lambda _2=2~z_2^2$, \\ $10~\xi_2^2+11~\xi_2+3\neq 0$\end{tabular}  & $-4 z_2$ & $-z_2-1$ &  $\theta_1$ & $-\frac{-3 \xi_{2}+8 \xi_{2} z_2+4 z_2}{2 \left(2 \xi_{2}+1\right)}$ & \begin{tabular}{@{}c@{}}Stable for \\ $0\leq z_2<\frac{3}{8}\land -\frac{1}{2}<\xi_{2}<-\frac{4 z_2}{8 z_2-3}$\end{tabular}  \\   
     \hline
    \parbox[c][1.8cm]{0.7cm}{$C$ }  & $1$ & $y_3$ & $0$ & $0$ & \begin{tabular}{@{}c@{}}$\lambda_3=0$,$6 y_3+1\neq 0$, \\ $\gamma_{3}=\frac{1}{12} \left(6 y_3-1\right)$, \\ $10~\xi_{3}^2+11~\xi_{3}+3\neq 0$\end{tabular} & $0$ & $-\frac{12 \left(\xi_{3}+1\right)}{5 \xi_{3}+3}$ & $\frac{3 \xi_{3}}{2 \left(2 \xi_{3}+1\right)}$ & $-1$ & \begin{center}
        \begin{tabular}{@{}c@{}}Stable for \\ $-\frac{1}{2}<\xi_{3}<0$\end{tabular}
    \end{center}  \\
     \hline
    \parbox[c][1.3cm]{0.7cm}{$D$ }  & $1$ & $0$ & $z_4$ & $0$ &  \begin{tabular}{@{}c@{}}$\gamma_{4}=0$,$z_4\neq 0$,$\lambda_4=2~z_4^2$, \\ $2~\xi_{4}+1\neq 0$\end{tabular} & $6$ & $\frac{69}{16}$ & $\frac{39}{17}$ & $\frac{1}{2}$ & \begin{center}
        Unstable
    \end{center} \\
     \hline
   \parbox[c][1.8cm]{0.7cm}{$E$ }  & $1$ & $y_5$ & $z_5$ & $0$ & \begin{tabular}{@{}c@{}}$2~z_5^2+3~z_5-18\neq 0$,$\gamma_{5}$,\\ $ 5~z_5+6\neq 0$, $\xi_{5}=-\frac{3 \left(z_5+2\right)}{5 z_5+6}$, \\ $\lambda _5=2~z_5^2,z_5+6\neq 0$\end{tabular} & $0$ & $-4z_5$ & $-z_5-1$ & $-\frac{4 z_5^2+15 z_5-18}{2 \left(z_9+6\right)}$ & \begin{center}
       \begin{tabular}{@{}c@{}}Stable for \\ $z_5>\frac{3}{8} \left(\sqrt{57}-5\right)$\end{tabular}
   \end{center} \\
 \hline
    \end{tabular}
\caption{Critical Points, Conditions for Existence, Eigenvalues and Stability Conditions.$(\gamma_{5}=\frac{-8 y_5 z_5^2-30 y_5 z_5+36 y_5-3 z_5-6}{-8 z_5^2-12 z_5+72})$ and $(\theta_1= -\frac{2 \left(6 \xi_{2}+5 \xi_{2} z_2+3 z_2+6\right)}{5 \xi_{2}+3})$}
    \label{table2}
\end{table}

\begin{table}[H]
    \centering 

    \begin{tabular}{c c c c c c c} 
    \hline\hline 
    \parbox[c][0.9cm]{1.3cm}{Critical Points
    }&  $q$ & $\omega_{tot}$ & $\omega_{eff}$ &$\Omega_m$ & $\Omega_r$ & $\Omega_{de}$ \\ [0.5ex] 
    \hline\hline 
    \parbox[c][1.3cm]{1.3cm}{$A$ } & $1$ & $\frac{1}{3}$ & - & $1-v_1$ & $v_1$ & $0$\\
    \parbox[c][1.3cm]{1.3cm}{$B$ } & $-z_2-1$ & $-\frac{2 z_2}{3}-1$ & $\frac{2 z_2}{3}-1$ & $0$ & $0$ & $1$\\
    \parbox[c][1.3cm]{1.3cm}{$C$ }  & $-1$ & $-1$ & $-1$ & $0$ & $0$ & $1$\\
    \parbox[c][1.3cm]{1.3cm}{$D$ }  & $-z_4-1$ & $-\frac{2 z_4}{3}-1$ & $\frac{7 \left(\xi _4+1\right) \left(2 z_4-3\right)}{3 \left(6 \xi _4+7\right)}$ &$1$  & $0$ & $0$ \\
   \parbox[c][1.3cm]{1.3cm}{$E$ }  & $-z_5-1$ & $-\frac{2 z_5}{3}-1$ & $\theta_2$ & $0$  & $0$ & $1$\\
 \hline
    \end{tabular}
    \caption{Values of different parameters Corresponding to different Critical points.$\left(\theta_2=\frac{324-324~z_5~(1+y_5)+9~z_{5}^2~(1+6y_5)+4~z_{5}^3~(7+6~y_5)}{3~(-72+6~y_5(6+z_5)~(6+5~z_5)+z_5(54+17z_5))}\right)$}
    \label{table3}
\end{table}

    \begin{table}[H]
    \centering 

    \begin{tabular}{c c c c c} 
    \hline\hline 
    \parbox[c][0.9cm]{1.3cm}{Critical Points
    }& {Acceleration equation} & {Scale factor (Power law solution)} & {Universe Phase} & Type of Expansion  \\ [0.5ex] 
    \hline\hline 
    \parbox[c][1.3cm]{1.3cm}{$A$ } & $\dot H=-2~H^2$ & $a(t)=t_0(2t+c_2)^\frac{1}{2}$ & {Radiation-dominated} & {Decelerating}\\
    \parbox[c][1.3cm]{1.3cm}{$B$ } & $\dot H=0$ & $a(t)=t_0e^{c_1t}$ & {de-Sitter phase} & {Accelerating} \\
    \parbox[c][1.3cm]{1.3cm}{$C$ }  & $\dot H=0$ & $a(t)=t_0e^{c_1t}$ & {de-Sitter phase} & {Accelerating}\\
    \parbox[c][1.3cm]{1.3cm}{$D$ }  & $\dot H=-\frac{3}{2}H^2$ & $a(t)=t_0(\frac{3}{2}t+c_3)^{\frac{2}{3}}$ & {Matter-dominated phase} & {Decelerating}\\
   \parbox[c][1.3cm]{1.3cm}{$E$ }  & $\dot H=0$ & $a(t)=t_0e^{c_1t}$ & {de-Sitter phase} & {Accelerating}\\
 \hline
    \end{tabular}
  
    \caption{Evolution equation and exact solutions corresponding to each Critical Point.}
    \label{table4}
\end{table}
    
  \begin{figure}[H]
    \centering
    \includegraphics[width=5.8cm]{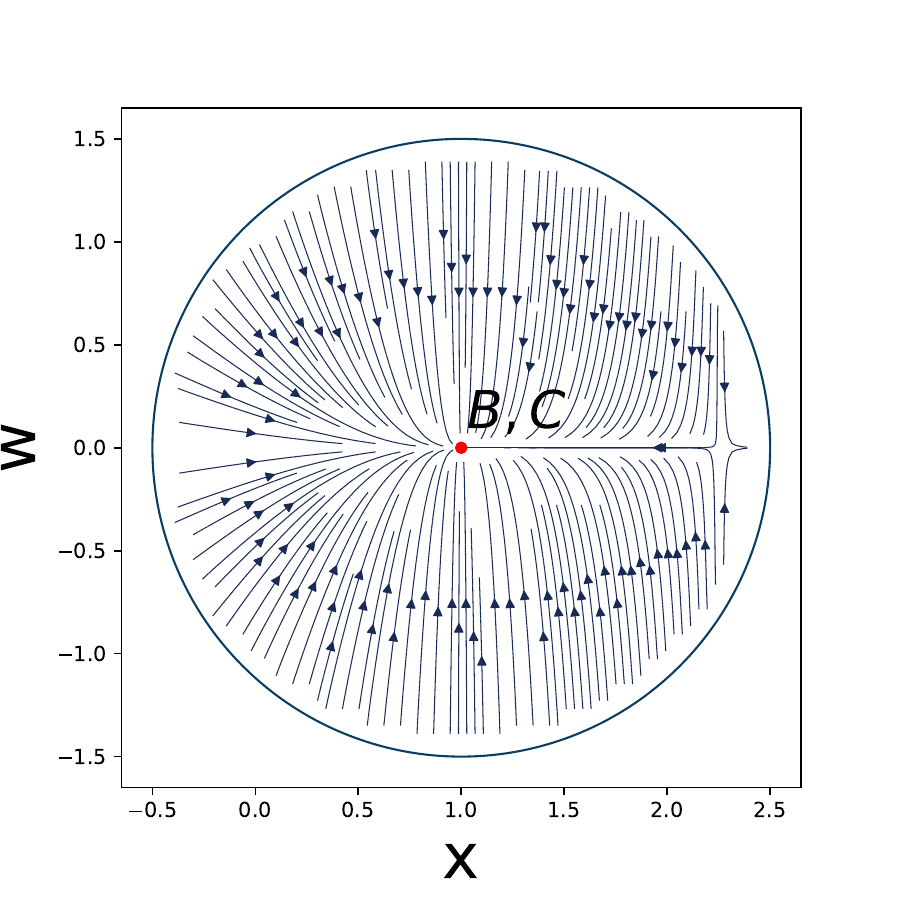}
    \includegraphics[width=5.8cm]{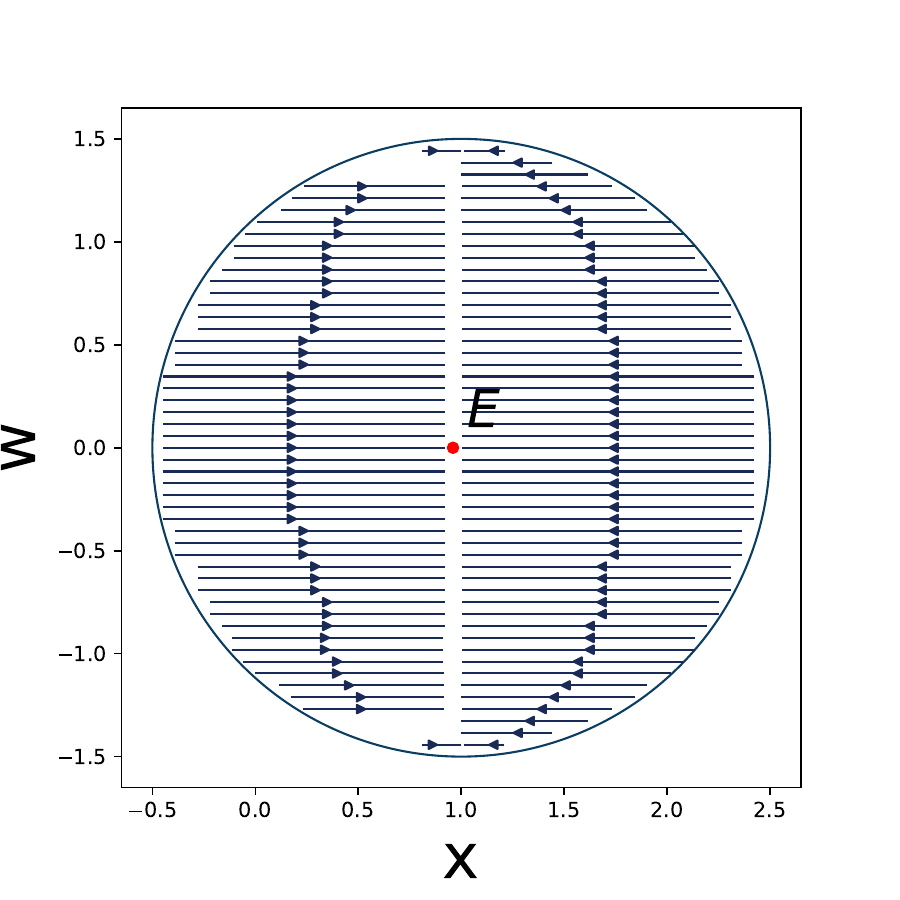}
    \includegraphics[width=5.8cm]{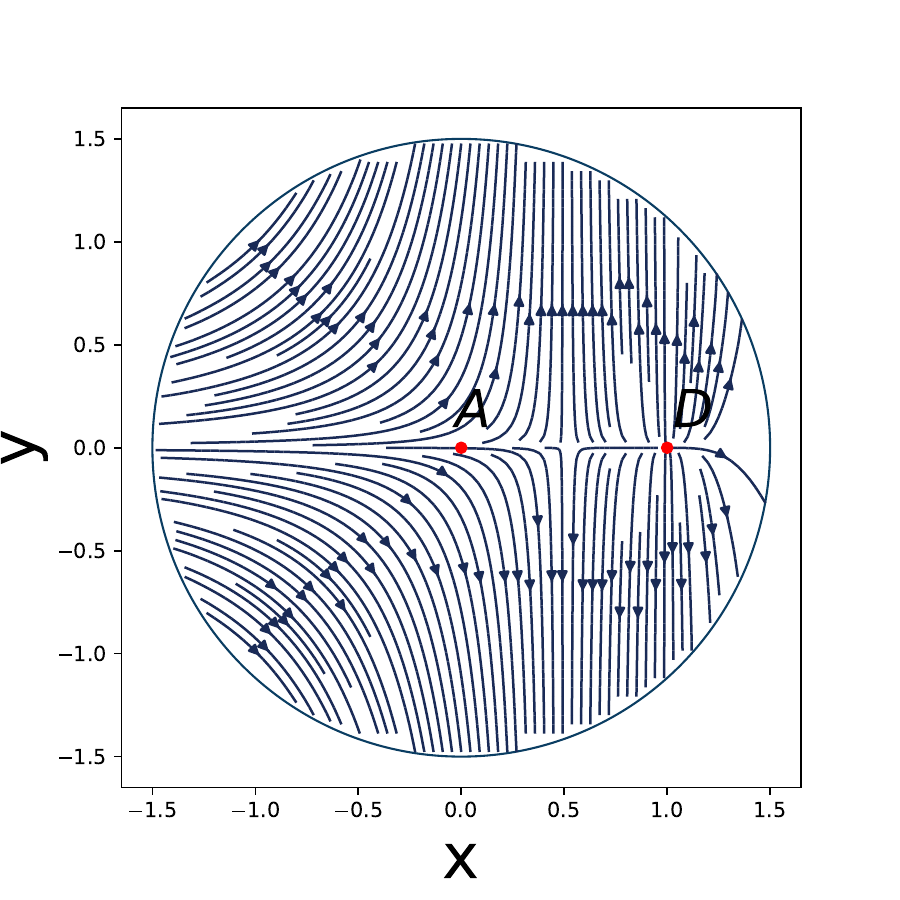}
    \caption{Phase portrait for the dynamical system.}
    \label{Fig.6}
\end{figure}

\begin{figure}[H]
    \centering
    \includegraphics[width=10cm]{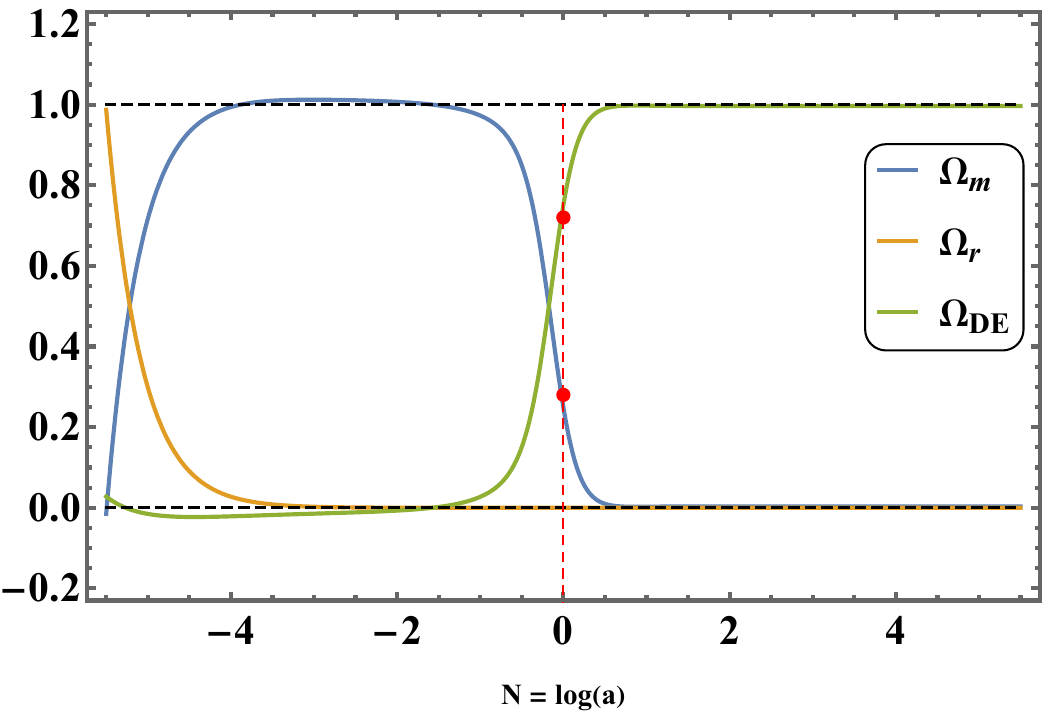}
     \includegraphics[width=15cm]{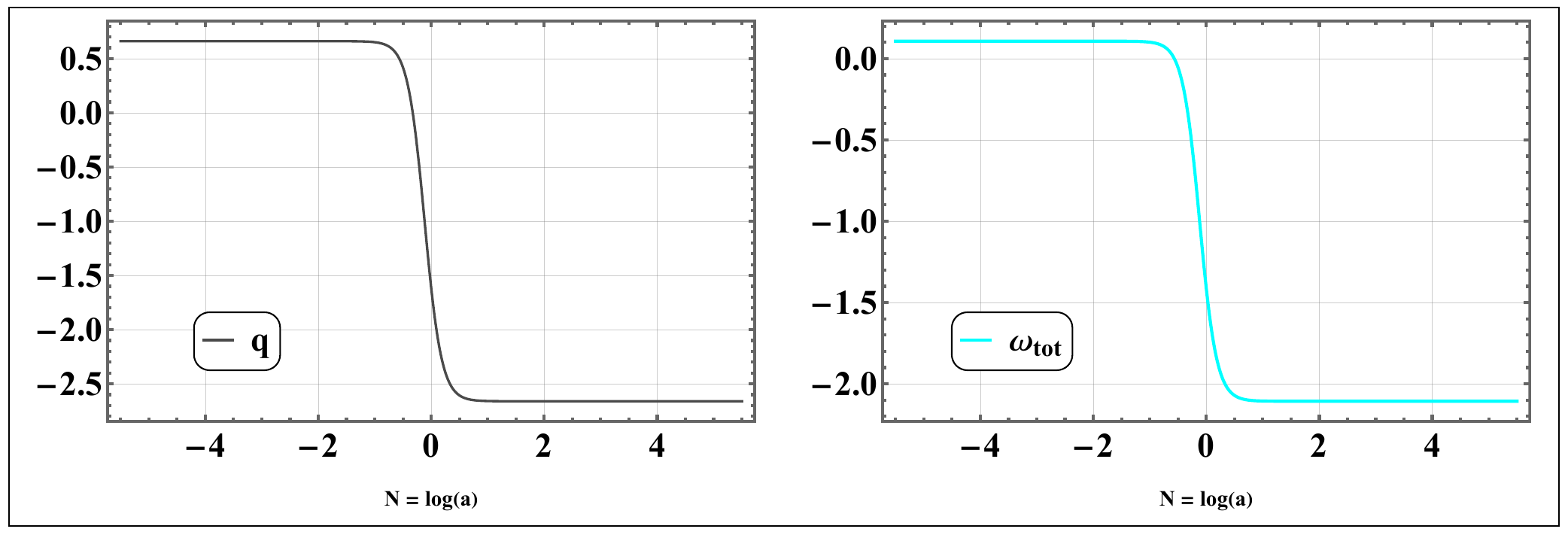}
    \caption{Evolution Plots for Density Parameter, Deceleration and EoS Parameter. The initial conditions: $x=0.6067$, $y=-0.048$, $z=0.62$ and $v=5.89\times10^{-7}$ for the system of differential Eq. \eqref{Eqn.43}.} 
    \label{Fig7}
\end{figure}

The details description of each of the critical points are listed below:

\begin{itemize}
    \item \textbf{Critical Point $A$: Radiation Dominated Phase}: At this critical point, we calculate the value of $\omega_{tot}=\frac{1}{3}$, which indicates that the Universe is filled by radiation-like fluid, meaning this point corresponds to the radiation-dominated phase of the universe. For that phase, we also calculate the value of the deceleration parameter, which gives $q=1$. The positive value shows that the universe was decelerating at that phase. Also, the values of density parameters are $\Omega_m=1-v_1$, $\Omega_r=v_1$ and $\Omega_{de}=0$, so the value of $v_1=1$ indicates total dominance of radiation. In FIG.\ref{Fig.6} [{\bf(Right Panel)}] point $A$ in the $xy$- plane, which always shows saddle behaviour trajectories that come toward the $y$-axis and then show repealing behaviour after applying conditions that satisfy the critical point for existence.

    \item \textbf{Critical Point $D$: Matter Dominated Phase}: This point is a repeller due to the positive eigenvalue. This point corresponds to the matter-dominated phase at the value of $\xi_4=-\frac{7}{6}$. For this, we calculate the density parameter, and it gives a value for $\Omega_m=1$ dominance of matter, and all other density parameters are $\Omega_r=0$ and $\Omega_{de}=0$. At $z_4=-\frac{3}{2}$, it gives the value of the total EoS parameter $\omega_{tot}=0$ and $q=\frac{1}{2}$, which indicates that the universe was decelerating at that phase. The phase space trajectory unstable behaviour in the $xy$- plane and eigenvalue indicates the instability of the point $D$ [FIG.\ref{Fig.6} {\bf(Right Panel)}].

    \item  \textbf{Critical Points $B$, $C$, $E$: De-Sitter Phase}: These points provide stable behaviour at $0\leq z_2<\frac{3}{8}\land -\frac{1}{2}<\xi_{2}<-\frac{4 z_2}{8 z_2-3}$, $-\frac{1}{2}<\xi_{3}<0$ and $z_5>\frac{3}{8} \left(\sqrt{57}-5\right)$ respectively for $B$, $C$ and $E$. For the point $C$, we calculate the value of $\omega_{tot}=-1$, which indicates that the asymptotic solution is the de-sitter universe. On the other hand, points $B$ and $E$ describe a family of points with $\omega_{tot}=-1-\frac{2 z}{3}$, which means the asymptotic solution provides a de-sitter universe for $z\geq-1$ and at $z=0$, the de-sitter universe recovered. Also, $z\geq-1$ gives the negative value of the deceleration parameter, indicating the universe is accelerating. Here, we calculate the density parameter that gives $\Omega_{de}$ for $E$ at point $y_5=-\frac{1}{6} $ and $ 2 z_5^2+3 z_5-18\neq 0$. A critical point, $C$, provides negative and zero eigenvalues for all free parameters. These eigenvalues associated with critical points are non-hyperbolic critical points. It has been mentioned that \cite{Coley_1999,aulbach_1984_1058}, the dimension of the set of eigenvalues for critical points is equal to the number of vanishing eigenvalues. This eigenvalue set is characterized by normal hyperbolicity, indicating stability of the corresponding critical point. So, for the critical point $E$ to make normal hyperbolic, it needs to satisfy the condition, $z_{5}>\frac{3}{8} \left(\sqrt{57}-5\right)$. Using this condition, we can fix $z_5$ and make $E$ normally a hyperbolic critical point showing stability. The phase space trajectory for critical points $B$ and $C$ shows attracting behaviour in the $xw$- plane [FIG.\ref{Fig.6}, {\bf(Left Panel)}]. For point $E$ [FIG.\ref{Fig.6}, {\bf(Middle Panel)}], phase space trajectory in the $xw$-plane shows attracting behaviour at $(1,0)$ after applying a particular condition on the remaining variable without violating the condition of existence and stability.
\end{itemize}
 Further, the evolutionary behaviour of the Universe can be analysed through the dynamical and geometrical parameters obtained in terms of dynamical variables. In FIG. \ref{Fig7} {\bf (Upper Panel)}, the evolutionary behaviour of density parameters of radiation, matter and de Sitter phase has been shown. It has been observed that, the early phase of the universe was dominated by radiation, then the matter-dominant era began. At present {\bf (vertical red line)}, $\Omega_r\approx0$, $\Omega_m\approx0.287$ and $\Omega_{de}\approx0.702$. It has been noticed that, DE is dominating at late times. In In FIG. \ref{Fig7} {\bf (Lower Panel)}, the evolution of deceleration and EoS parameters has been shown. The deceleration parameter shows that in early phase, it is showing decelerating behaviour whereas at late times, it is accelerating. Also, the behaviour of the EoS parameter shows that the universe is in the accelerated expansion phase in late time.   

\section{Results and Discussions}\label{Sec:6} 
The accelerating cosmological model constrained with the cosmological observations has been presented in Weyl type $f(Q,T)$ gravity. The parametric form of Hubble parameter has been introduced to find the constrained range of the parameters. The evolution of Hubble rate has been illustrated with $CC$, $Pantheon^+$, and $BAO$ datasets $\left[FIG. \ref{fig:II}\right]$. At low redshift, the projected values of $H(z)$ closely resembles to that of GR with the cosmological constant; however at higher redshifts, significant distinction has been observed with that of   $\Lambda$CDM model. The transient behaviour of deceleration parameter shows the accelerating behaviour at late epochs of evolution. The behaviour of dynamical parameters also support the accelerating behaviour of the model with the considered form of $f(Q,T)$. Further the violation of SEC and non-violation of NEC and DEC supports the behaviour in the context of modified gravity.\\

Further we have studied the dynamical system analysis of the model and analysed each critical points individually. We have also connected the cosmological parameters in terms of dynamical variables. We conducted the dynamic analysis within the phase space, identifying the critical points along with their cosmological attributes and stability criteria. The critical points and their existence condition, are outlined in TABLE \ref{table2}, while the formulations for the cosmological parameters pertaining to each critical point are detailed in TABLE \ref{table4}. Moreover, TABLE \ref{table3} illustrates the conditions necessary for stability, and acceleration. We noticed that, within the framework of dynamical systems approach, the presence of unstable critical points represents the decelerated expansion during early time such as the radiation and matter-dominated epochs. Whereas, the  stable critical points exhibited accelerated expansion at late times i.e in the de Sitter phase. We can conclude that the model exhibits accelerating behaviour in both the approaches such as with the use of cosmological datasets and by using dynamical system analysis.

\section*{Acknowledgement} RB acknowledges the financial support provided by University Grants Commission (UGC) through Junior Research Fellowship UGC-Ref. No.: 211610028858 to carry out the research work. BM acknowledges the support of IUCAA, Pune (India) through the visiting associateship program.

\section*{References}
\bibliographystyle{utphys}
\bibliography{references}

\end{document}